\documentstyle[12pt,epsf,epsfig]{article} 
\input epsf

\newcommand{\frat}[2]{\frac{\textstyle #1}{\textstyle #2}}
\newcommand{\dmn}[2]{\mbox{$#1\!\cdot\! 10^{#2}\,$}}
\newcommand{\grpicture}[1]
{\epsfxsize=200pt 
    \hspace{5cm} \parbox{200pt}{\epsfbox{#1.ps}} \vspace{5mm}
}

\begin{document}
\begin{center}
{\Large 
\bf On the origin of current quark mass\\ within nonperturbative QCD}\\
\vspace{0.5cm}
{S.V. Molodtsov, A.M. Snigirev$^\dagger$, G.M. Zinovjev$^\ddagger$}\\
\vspace{0.5cm} 
 {\small\it State Research Center,
Institute of Theoretical and Experimental Physics,
117259, Moscow, Russia\\ 
$^\dagger$Skobeltsyn Institute of Nuclear Physics, Moscow State
University,
119899, Moscow, Russia\\
$^\ddagger$Fakult\"at f\"ur Physik, Universit\"at Bielefeld,
D-33501, Bielefeld, Germany\\
$^\ddagger$Bogolyubov Institute for Theoretical Physics,\\
National Academy of Sciences of Ukraine, 
UA-03143, Kiev, Ukraine} 
\end{center}
\vspace{0.5cm}
\begin{abstract}
We develop the theory of fermion induced phononlike excitations of
the instanton liquid. It suggests a mechanism of current quark mass 
generation which is easily understandable by calculating the corresponding 
functional integral in the tadpole approximation. We systematically 
study the quark condensate excitations influenced by the phononlike 
modes and rederive the relation Gell-Mann--Oakes--Renner with
realistic pion mass. The picture of $\sigma$-meson as being mixed 
with the soft scalar glueball-like excitation is discussed.  
\end{abstract}
\vspace{0.5cm}
\\
\\ 
\section*{I. Introduction}
Nowadays, there are no doubts the model of QCD vacuum as the instanton liquid 
(IL) 
is the most practical instrument on the chiral scale of QCD. It provides, 
as the
lattice calculations recently confirmed, not only the theoretical 
background for describing spontaneous chiral symmetry breaking (SCSB) but
is mostly powerful in the phenomenology of the QCD vacuum and in the physics
of light quarks while considered to propagate by zero modes arising
from instantons. The origin of gluon and chiral condensates turns out in
this picture easily understandable and both are quantitatively calculated
getting very realistic values defined by $\Lambda_{QCD}$ and parameters of
instanton and anti-instanton ensemble, for example, 
$-i~\langle\psi^\dagger\psi\rangle\sim -(250~MeV)^3$. Moreover, the scale
for dynamical quark masses, $M\sim 350~MeV$, naturally appears and pion
decay constant, $f_\pi\sim 100~MeV$, is then transparently calculated.

Another significant advantage of this approach is that the initial formulation
starts basically from the first principles and subsequent approximations
being well grounded and reliably controlled are plugged in \cite{1},
\cite{2}. It
becomes clear especially in recent years when the impressive progress has 
been reached in understanding the instanton physics on the lattice
\cite{Negele}. Further we are summarizing several things we have learned 
thinking of the IL theory and trying to answer the challenging questions.

Let us start on that stage of the IL approach when its generating functional 
has already been taken as factorized one into two factors
$$ 
{\cal Z}~=~{\cal Z}_g~\cdot~{\cal Z}_\psi~,
$$
where eventually ${\cal Z}_{g}$ provides nontrivial gluon condensate and
the fermion part ${\cal Z}_{q}$ is responsible to describe the chiral
condensate in instanton medium and its excitations. It is usually
supposed the functional integral of ${\cal Z}_{g}$ is saturated by 
the superposition of the pseudo-particle (PP) fields which are the
Euclidean solutions of the Yang-Mills equations called the 
(anti-)instantons
\begin{equation}
\label{2} 
A_\mu(x)=\sum_{i=1}^N A_\mu(x;\gamma_i)~.
\end{equation}
Here $A_\mu(x;\gamma_i)$ denotes the field of a single (anti-)instanton
in singular gauge with $4N_c$ (for the $SU(N_c)$ group) coordinates 
$\gamma=(\rho,~z,~U)$ of size $\rho$ centred at the coordinate $z$
and colour orientation defined by the matrix $U$. The nontrivial bloc of
corresponding $N_{c} \times N_{c}$ matrices of PP is a part of potential
\begin{equation}
\label{3}
A_\mu(x;\gamma)=\frat{\bar \eta_{a\mu\nu}}{g}
\frat{y_\nu}{y^2}\frat{\rho^2}{y^2+\rho^2}~U^\dagger \tau_a~U,~~y=x-z~,
~a=1,2,3~,
\end{equation}
where $\tau_a$ are the Pauli matrices, $\eta$ is the 't Hooft
symbol \cite{3}, $g$ is the coupling constant
and for anti-instanton $\bar \eta \to \eta$. For the sake
of simplicity we do not introduce the distinct symbols for instanton
($N_{+}$) and anti-instanton ($N_{-}$) and consider topologically neutral
IL with $N_+=N_-=N/2$. Utilizing the variational principle
the following estimate of ${\cal Z}_{g}$ was found \cite{2} 
$$
{\cal Z}_g~\simeq~e^{-\langle S\rangle}
$$
with the action of IL defined by the following additive functional
{\footnote{
In fact, the additive property results from the supposed homogeneity of
vacuum wave function in metric space. Eq. (\ref{s}) looks like a formula of
classical physics although it describes the ground state of quantum
instanton ensemble. Intuitively clear, this definition will be still
valid even when the wave function is nonhomogeneous with the
nonuniformity scale essentially exceeding average instanton size or, 
precisely speaking, being larger (or of the order) than average size of 
characteristic 
saturating field configuration. Then each instanton liquid element of 
such a distinctive size will provide a partial contribution depending 
on the current state of IL, see next Section.}}
\begin{equation}
\label{s}
\langle S\rangle=\int d z \int d\rho~n(\rho)~s(\rho)~.
\end{equation}
The integration should be performed over the IL volume $V$
along with averaging the action per one instanton  
\begin{equation}
\label{si}
s(\rho)=\beta(\rho)+5 \ln(\Lambda\rho)-\ln \widetilde \beta^{2N_c}
+\beta \xi^2\rho^2\int d\rho_1~n(\rho_1)\rho_1^{2},
\end{equation}
weighted with instanton size distribution function
\begin{equation}
\label{nrho}
n(\rho)=C~e^{-s(\rho)}=C~\rho^{-5} \widetilde\beta^{2 N_c} 
e^{-\beta(\rho)-\nu \rho^2/\overline{\rho^2}}~,
\end{equation}
$$
\nu=\frat{b-4}{2},~b=\frat{11~N_c-2~N_f}{3}~,~~
\left(\overline{\rho^2}\right)^2=
\frat{\nu}{\beta \xi^2 n},$$
where
$\overline{\rho^2}=\int d\rho~\rho^2~n(\rho)/n,~n=\int d\rho~n(\rho)=N/V$ 
and $N_f$ is the number of flavours.
The constant $C$ is defined by the variational maximum principle in the
selfconsistent way and $\beta(\rho)=\frat{8\pi^2}{g^2}=
-\ln C_{N_c}-b \ln(\Lambda \rho)
~(\Lambda=\Lambda_{\overline{MS}}=0.92 \Lambda_{P.V.})$ 
with constant $C_{N_c}$ depending 
on the renormalization scheme, in particular, here 
$C_{N_c}\approx\frat{4.66~\exp(-1.68 N_c)}
{\pi^2 (N_c-1)!(N_c-2)!}$. The parameters $\beta=\beta(\bar\rho)$ and 
$\widetilde \beta=\beta+\ln C_{N_c}$ are 
fixed at the characteristic scale $\bar\rho$ (an average instanton size).
The constant $\xi^2=\frat{27}{4}\frat{N_c}{N_c^{2}-1} \pi^2$ 
characterizes, in a sense, the PP interaction and Eqs. (\ref{s}),(\ref{si}) 
and (\ref{nrho}) describe the equilibrium state of IL.
The minor modification of variational maximum principle (see 
Appendix) leads to the explicit form of the mean instanton size 
$\bar\rho\Lambda=\exp\left\{-\frat{2N_c}{2\nu-1}\right\}$ and, therefore, 
to the direct definition of the IL parameters unlike the  
conventional variational principle \cite{2} which allows one to extract
those parameters solving numerically the transcendental equation only.

The quark fields are considered to be {\it influenced} by the certain
stochastic ensemble of PPs, Eq. (\ref{2}), while calculating the quark 
determinant   
$$
{\cal Z}_\psi~\simeq~\int D\psi^\dagger D\psi~~\langle\langle~
e^{S(\psi,\psi^\dagger,A)}~\rangle\rangle_A~.
$$
Besides, dealing with dilute IL (small characteristic  
packing fraction parameter $n~\bar\rho^4$) one neglects the correlations 
between PPs and utilizes the approximation of $N_c\to\infty$ where the planar 
diagrams only survive. In addition the fermion field action is 
approached by the zero modes what means the quark Green function is 
considered as the superposition of the free Green function 
$S_0=(-i\hat\partial)^{-1}$
and fermion zero modes $\Phi_\pm(x-z)$ which are the solutions of the Dirac 
equation $i(\hat D(A_\pm)+m)~\Phi_\pm=0$ in the field of (anti-)instanton 
$A_\pm$ centred at $z$, i.e. 
\begin{equation}
\label{gf}
S_\pm=S_0-\frat{\Phi_\pm\Phi_{\pm}^\dagger}{i~m}~,
\end{equation} 
here $m$ is the current quark mass and the quark zero mode possesses the 
following analytic form
$$
\left[\Phi_\pm(x)\right]_{ic}=\frat{\rho}{\sqrt{2}\pi |x|(x^2+\rho^2)^{3/2}}
\left[\hat x~\frat{1\pm\gamma_5}{2}\right]_{ij}\varepsilon_{jd}
~U_{dc}
$$
with the colour $c,d$ and the Lorentz $i,j$ indices and the antisymmetric 
tensor $\varepsilon$. 
In fact, there exists the singular term in the limit $m\to0$ but it 
is selconsistently fixed by the saddle point calculation of quark determinant
${\cal Z}_\psi$. In spite of the fact the exact Green function
(at $m\neq0$) including the terms of the whole series is well known 
\cite{car} and, moreover, the Green function of instanton molecule 
has been also established \cite{lee}, the simple zero mode approximation 
(\ref{gf}) is still the most practical in the concrete evaluations. 
In particular, at $N_f=1$ the quark determinant reads \cite{2}
\begin{equation}
\label{6}
{\cal Z}_\psi~\simeq~\int D\psi^\dagger D\psi ~ 
\exp\left\{\int d x~\psi^\dagger(x)~i\hat\partial\psi(x)\right\}~
\left(\frat{Y^+}{VM}\right)^{N_+}~
\left(\frat{Y^-}{VM}\right)^{N_-}~,
\end{equation}
$$Y^{\pm}=i\int dz~dU~d\rho~n(\rho)/n
\int dxdy~\psi^\dagger(x)~i\hat\partial_x \Phi_\pm(x-z)~
\Phi^\dagger_{\pm}(y-z)~i\hat\partial_y~\psi(y)~, 
$$
where the factor $M$ makes the result dimensionless and is also fixed by the
saddle point calculation. Amazingly this relatively crude
approximation turns out so fruitful to develop (even quantitatively!) the  
low energy phenomenology of light quarks. The generating functional beyond
the chiral limit was obtained in Ref. \cite{mu}.

Thus, the IL approach at the present stage of its development looks very 
indicative, well theoretically grounded and reasonably adjusted 
phenomenologically. The proper form of generating functional obtained
and its verisimilar parameter dependence indicated provide enough predictive
power and justify, hence, the approximations made. It dictates
the improvements to be done within the approach but, on
the other hand, calculating some corrections has clearly no serious 
prospect. Taking this message as a guiding one we are going to demonstrate
that amplifying the approach with an {\it inverse influence} of quarks upon 
the
instanton ensemble which is intuitively small effect leads, however, to
rather unobvious important conclusions.

We describe this influence (not getting beyond IL and SCSB approach)
as a small variation of instanton liquid parameters $\delta n$ and 
$\delta \rho$ around their equilibrium values of $n$ and 
$\bar\rho$ being in full analogy with the description of chiral condensate
excitations. Indeed, the result of nontrivial calculation of the functional
integral (treating substantially from physical view point the zero quark 
modes in the fermion determinant) comes to 'encoding' the IL state with
just those two parameters. Moreover, the IL density 
appears in the approach via the packing fraction parameter $n\bar\rho^4$
only (clear from dimensional analysis) what means one independent
parameter existing in practice. It is just the instanton size. The
analysis of the {\it quark and IL interaction} is addressed in this paper
developing
our idea \cite{we} of phononlike excitations of IL resulting from the
adiabatic changes of the instanton size. Thus, we will suppose the
{\it inverse quark influence} should be described by these deformable 
(anti-)instanton configurations which are the 
field configurations Eq. (\ref{3}) characterized by the size $\rho$
depending on $x$ and $z$, i.e. $\rho\to\rho(x,z)$.

The paper is organized as follows: in Section II we discuss the
modification of quark determinant when the functional integral is
saturated by the deformable modes at the minimal number of flavours. 
Then in Section III we develop the approximate calculation (tadpole 
approximation) which is based on the saddle point method and the 
corresponding iteration procedure. Section IV is devoted to the 
generalization for the multiflavour picture. The meson excitations of 
quark condensate and calculation of the  Gell-Mann--Oakes--Renner relation
when it is provided by the mechanism of 
quark current mass generation related to the phononlike excitations of 
IL are analyzed in Section V. The paper includes also Appendix where the
fault finding reader gets a chance to control the explicit formulae of the 
IL parameters 
and to improve our calculations if is able.
 
\section*{II. Supplementing phononlike excitations}
Apparently, the gist of what we discuss here could be
illuminated in the following way.
Saddle point method calculation of the functional integral implies
the treatment of the action extremals which are the solutions of classical
field equations. For the case in hands the action $S[A,\psi^\dagger,\psi]$
is constructed including the gluon fields $A_{\mu}$, (anti-)quarks
$\psi^\dagger,\psi$ and extremals, which are given by the solutions of
the consistent system of the Yang-Mills and Dirac equations. As a trial
configuration in the IL theory the superposition of (anti-)instantons
which is the approximate solution of the Yang-Mills equations (with no
reverse influence of the quark fields) and an external field for the Dirac
equation simultaneously is considered. We ~believe it is reasonable to
utilize the deformable (crampled) (anti-)instantons $A_{\pm}(x;\gamma(x))$
as the saturating configurations. They just admit of varying the
parameters $\gamma(x)$ of the ~Yang-Mills sector of the initial
consistent system in 
order to describe the influence of quark fields in the appropriate 
variables for the quark determinant.

Taking the action in the form 
$S[A_{\pm}(x,\gamma(x)),\psi^\dagger,\psi]$ we would receive the corresponding
variational equation for the deformation field $\gamma(x)$ which 
approaches most optimally (as to the action extremum) PP
at nonzero quark fields. In the field theory, for example, the monopole
scattering \cite{man}, the Abrikosov vortices scattering \cite{serg} are
treated in a similar way. However, for IL we avoid the
difficulties which come with solving the variational equations if we 
consider the long-length wave excitations only with the wave length 
$\lambda$ much
larger the characteristic instanton size $\bar\rho$. Indeed, it looks
possible because we are searching the kinetic energy of the deformation
fields 
{\footnote{Then we are allowed to take the slowly changing deformation field
beyond the integral while calculating the action of deformed instanton.}} 
(the one particle contributions) 
and consider the pair interaction which develops
the contact interaction form being calculated in the adiabatic regime
\cite{we}. 

Let us remind first of all that deriving Eq. (\ref{s}) we should average
over the instanton positions in a metric space. Clearly, the characteristic
size of the domain $L$ which has to be taken into account should exceed the 
mean instanton size $\bar\rho$. But at the same time it should
not be too large
because the far ranged elements of IL are not 'causally' dependent. The
ensemble wave function is expected to be homogeneous (every PP contributes
to the functional integral being weighted with a factor proportional to
$\sim 1/V,~V=L^4$) on this scale. The characteristic configuration which
saturates the functional integral is taken as the superposition Eq. (\ref{2})
with $N$ of PP in the volume $V$. It is easy to understand that because of
an additivity of the functional Eq. (\ref{s}) describes properly even
non-equilibrium states of IL when the distribution function $n(\rho)$ 
does not coincide with the vacuum one and, moreover, allows us to
generalize it for the non-homogeneous liquid when the size of the
homogeneity obeys the obvious requirement $\lambda\ge L>\bar\rho$. 
Besides, we should deal with the weak (comparing to the instanton 
$G_{\mu\nu}$) fields
$g_{\mu\nu}$ $~(g_{\mu\nu}\ll G_{\mu\nu})$ and long-length wave (on the scale 
of the instanton size $\bar\rho$) perturbations 
$\left|\frat{\partial\rho(x,z)}{\partial x}\right|\ll O(1)$
when the ensemble saturating the functional integral is very close to the
vacuum (instanton) one.
Then this smoothness or the adiabatic change of instanton size practically
dictates another essential simplification defining everywhere the field
in the center of instanton  
$\frat{\partial\rho(x,z)}{\partial x}\sim
\left.\frat{\partial\rho(x,z)}{\partial x}\right|_{x=z}$ 
as a characteristic deformation field and
the exact instanton definition in the singular gauge,
$A^{a}_\mu(x,z)=-\frat{\bar\eta_{a\mu\nu}}{g}\frat{\partial}
{\partial x_\nu}\ln\left(1+\frat{\rho^2}{y^2}\right)$,
leads to the following correction to the potential
\begin{equation}
\label{pop}
a^{a}_\mu(x)=\Phi^{a}_{\mu\nu}(x,z)~\frat{\partial \rho}{\partial x_\nu}~,
\end{equation}
where $\Phi^{a}_{\mu\nu}(x,z)=-\frat{\bar\eta_{a\mu\nu}}{g}
\frat{2\rho(x,z)}{y^2+\rho^2(x,z)}$.
Keeping ourselves within the precision accepted here we could take 
$\rho(x,z)\simeq\bar\rho$, i.e.
$\Phi^a_{\mu\nu}(x,z)=\Phi^a_{\mu\nu}(x-z)$.
On the  other hand the most general form of the correction
to the instanton field influenced by the quarks might be calculated
if the gluon Green function in the instanton medium is known 
\begin{equation}
\label{gGF}
a^a_{\mu}(x,z)=\int d\xi~D^{ab}_{\mu\nu}(x-z,\xi-z)
~J^{b}_\nu(\xi-z_\psi)~,
\end{equation}
where $J^{b}_\nu$ is the current of external quark source, 
$z_\psi$ belongs to the region of long-length wave disturbation and, 
at last, 
$D^{ab}_{\mu\nu}$ is the Green function of PP in the instanton medium. 
In fact, this function is not well defined \cite{car} but,  
seems, for the case in hands we could develop the selfconsistent way  
to calculate the regularized Green function. The nonsingular 
propagator behaviour in the soft momentum region is defined
by the mass gap of phononlike excitations. Fortunately,  the exact form
of the Green function occurs unessential here
(we are planning to return to the problem of regularized Green function 
calculation in the forthcoming publication).
In the coordinate space it is peaked
around the averaged PP size being in nonperturbative regime
and, hence, integral Eq. (\ref{gGF}) is appraised to be
\begin{equation}
\label{est}
a^a_{\mu}(x,z)\simeq F^{ab}_{\mu\nu}(x-z)~J^b_{\nu}(z-z_\psi)~.
\end{equation}
Appealing now to Eq. (\ref{pop}) we are capable to get immediately for 
$\rho_\nu$ the following equation
$$\Phi^a_{\mu\nu}(x-z)~\frat{\partial \rho(x,z)}{\partial x_\nu}=
F^{ab}_{\mu\nu}(x-z)~J^b_{\nu}(z-z_\psi)~.
$$
In fact, the current $J$ might be taken ~constant in the long-length wave 
approximation. Then neglecting the gradients
the following change appears to be justified everywhere
$\left |\frat{\partial\rho(x,z)}{\partial x}\right |\sim
\left |\frat{\partial\rho(z)}{\partial z}\right |$
since there is no other fields in the
problem at all (in the adiabatic approximation). 
The deformable mode contribution to the functional integral
(when the corrections coming from the deformation fields of PP are
absorbed) may be estimated as \cite{we}
$$
\langle S\rangle\simeq\int d z \int d\rho~n(\rho)
~\left\{~\frat{\kappa}{2}~\left(\frat{\partial \rho}{\partial z}\right)^2
+s(\rho)\right\}~,
$$
where $\kappa$ is the kinetic coefficient being derived within
the quasiclassical approach. Our estimate of it gives the value of a few 
instanton actions $\kappa\sim c~\beta$ with the coefficient 
$c\sim 1.5\div 6$ depending quantitatively on the ansatz
supposed for the saturating configurations.  Although this estimate
is not much meaningful because there is no the vital 
$\kappa$ dependence eventually (becomes shortly clear). 
Thus, this coefficient 
should be fixed on a characteristic scale, for example 
$\kappa\sim\kappa(\bar\rho)$ if we are planning not to be beyond the 
precision peculiar to the approach. Actually, it means adding the 
small contribution of kinetic energy type to the action
per one instanton only. Such a term results from the scalar field
of deformations and affects negligibly the pre-exponential factors
of the functional integral. In one's turn pre-exponential factors
do the negligible influence on the kinetic term as well
\cite{we}. If we strive for to be within the approximation we should 
retain the small terms of the second order in deviation from the 
point of action minimum only $\left.\frat{d s(\rho)}{d\rho}
\right|_{\rho=\rho_c}=0$ 
supposing approximately
\begin{equation}
\label{dec} 
s(\rho)\simeq s(\bar\rho)+\frat{s^{(2)}(\bar\rho)}{2}~\varphi^2~,
\end{equation} 
where 
$s^{(2)}(\bar\rho)\simeq\left.\frat{d^2 s(\rho)}{d\rho^2}
\right|_{\rho_c}=\frat{4\nu}{\overline{\rho^2}}$ 
and the scalar field
$\varphi=\delta\rho=\rho-\rho_c\simeq\rho-\bar\rho$
is the field of deviations from the equilibrium value of  
$\rho_c=\bar\rho~\left(1-\frat{1}{2\nu}\right)^{1/2}\simeq\bar\rho$. 
Consequently, the deformation field is described by the following 
Lagrangian density 
$$
{\cal L}=\frat{n\kappa}{2}
\left\{~\left(\frat{\partial \varphi}{\partial z}\right)^2+
M^2\varphi^2\right\}
$$
with the mass gap of the phononlike excitations
$$
M^2=\frat{s^{(2)}(\bar\rho)}{\kappa}=
\frat{4\nu}{\kappa \overline{\rho^2}}
$$
which is estimated for IL with $N_{c}=3$, for example, in the   
quenched approximation to be
$$M\approx 1.21~\Lambda~~$$   
if 
$c=4,~\bar\rho~\Lambda\approx 0.37,~\beta\approx 17.5,~n~\Lambda^{-4}
\approx 0.44$ 
(for the details see the tables of Appendix).  
The deformation fields $\varphi$ (with corresponding Jacobian)
contribute to the generating functional
on the same footing as the quark fields and it looks like
$$
{\cal Z}^{'}_g\sim\int D\varphi~
\left|\frat{\delta A}{\delta \varphi}\right|~
\exp\left\{-\frat{n\kappa}{2}\int dz
\left[~\left( \frat{\partial \varphi}
{\partial z}\right)^2+M^2\varphi^2\right]\right\}
$$
in full analogy with the fields $\psi^\dagger,~\psi$ entering with the
functional measure $D\psi^\dagger,~D\psi$. 
Actually  the Jacobian contribution
{\footnote{Generally, the deformation field $\rho_\nu$ and
integration variable $a^a_{\mu}$ (\ref{pop}) are related via the rotation 
matrix: $\Omega_{ab}\Phi^b_{\mu\nu}(x-z) \rho_\nu=\widetilde a^{a}_\mu$
and in the long-length wave approximation $\Phi$ might be constant 
 $\Phi^b_{\mu\nu}(0)~$($x\sim z$). 
With the rotation matrix spanning the colour field $a_\mu=\Omega^{-1}
\widetilde a_\mu$
on the fixed axis (on the z axis for SU(2) group,
for instance) we can conclude that the vectors 
$a^z_{\mu}$ and $\rho_\nu$ are, in fact, in one to one correspondence
(of course, being within one loop approximation and up to this unessential 
colour rotation). Thus, the Jacobian occurs an unessential constant. 
}}
should be omitted in what follows as discussed above.

Analyzing the modifications arising now in the quark determinant 
${\cal Z}_\psi$ we take into account the variation of fermion 
zero modes resulting from the instanton size perturbed 
$$\Phi_\pm(x-z,\rho)\simeq\Phi_\pm(x-z,\rho_c)+
\Phi_{\pm}^{(1)}(x-z,\rho_c)~\delta\rho(x,z)~,$$
where 
$\Phi^{(1)}_\pm(u,\rho_c)=\left.\frat{\partial\Phi_\pm(u,\rho)}
{\partial\rho}\right|_{\rho=\rho_c}$ 
and because of the adiabaticity  it is
valid $\delta\rho(x,z)\simeq$ $~\delta\rho(z,z)=\varphi(z)$. The
additional contributions of scalar fields generate the corresponding
corrections in the factors of the kernels $Y^{\pm}$ of Eq. (\ref{6}) 
which are treated in the linear approximation in $\varphi$, i.e.
\begin{eqnarray}
\label{8}
i\hat\partial_x \Phi_\pm(x-z,\rho)~
\Phi^\dagger_{\pm}(y-z,\rho)~i\hat\partial_y\simeq
\Gamma_\pm(x,y,z,\rho_c)+\Gamma^{(1)}_\pm(x,y,z,\rho_c)~\varphi(z)~,
\end{eqnarray}
here we introduced the notations
$$\Gamma_\pm(x,y,z,\rho_c)=i\hat\partial_x \Phi_\pm(x-z,\rho_c)
\Phi^\dagger_{\pm}(y-z,\rho_c)i\hat\partial_y~,$$
$$\Gamma^{(1)}_\pm(x,y,z,\rho_c)=
i\hat\partial_x \Phi^{(1)}_\pm(x-z,\rho_c)
\Phi^\dagger_{\pm}(y-z,\rho_c)~i\hat\partial_y+
i\hat\partial_x \Phi_\pm(x-z,\rho_c)~\Phi^{\dagger (1)}_{\pm}(y-z,\rho_c)~
i\hat\partial_y$$
(the gradients of scalar field $\varphi$ are negligible according to 
the adiabaticity assumption again). It is a simple matter to verify that 
the right hand side of  Eq. (\ref{8}), being integrated over $dzdU$, 
generates the following kernel (in the momentum space)
\begin{equation}
\label{8a} 
\frat{1}{N_c}\left[(2\pi)^4~\delta(k-l)~\gamma_0(k,k)+
\gamma_1(k,l)~\varphi(k-l)\right]\nonumber
\end{equation}
with $\gamma_0(k,k)=G^2(k)~,~G(k)=2\pi\rho_cF(k\rho_c/2)~,
~\gamma_1(k,l)=G(k)G'(l)+G'(k)G(l)~,~\\
G'(k)=\left.\frat{d G(k)}{d\rho}\right|_{\rho=\rho_c},~
F(x)=2x~[I_0(x)K_1(x)-I_1(x)K_0(x)]-2~I_1(x)K_1(x)$,
where $I_i,~K_i~(i=0,1)$ are the modified Bessel functions.

In fact, the functional integral of Eq. (\ref{6}) including the
phononlike component may be exponentiated in the momentum space 
{\footnote{In the metric space we have the nonlocal 
Lagrangian of the phononlike deformations  
$\varphi(z)$ interacting with the quark fields $\psi^\dagger,~\psi$, i.e.
\begin{eqnarray}
\label{ms}
{\cal L}&=&\int dx~\psi^\dagger(x)i\hat\partial_x \psi(x)-
\int dz~ \frat{n\kappa}{2}
 \left\{\left(\frat{\partial \varphi}{\partial z}\right)^2
+M^2 \varphi^2(z)\right\}+\nonumber\\
&+&
\frat{i\lambda_\pm}{N_c}\int dxdydz~dU~
\psi^\dagger(x)\{\Gamma_\pm(x,y,z,\rho_c)+
\Gamma^{(1)}_\pm(x,y,z,\rho_c)~\varphi(z)\}\psi(y)~.
\nonumber
\end{eqnarray}
The physical meaning of the basic phenomenon behind this Lagrangian seems
pretty transparent. The propagation of quark fields through the instanton 
medium is accompanied by the IL disturbance (the analogy with well known 
polaron problem embarrasses us strongly in this point).}} 
with the auxiliary integration over the 
$\lambda$-parameter (see, for example \cite{2}) 
\begin{eqnarray}
\label{9}
&&{\cal Z}_\psi\simeq\int \frat{d\lambda}{2\pi}~
\exp\left\{N\ln\left(\frat{N}{i\lambda VM}\right)-N\right\}\times\nonumber\\
&&\times\int D\psi^\dagger D\psi~ 
\exp\left\{\int \frat{dkdl}{(2\pi)^8}~\psi^\dagger(k)
\left[(2\pi)^4\delta(k-l)
\left(-\hat k+\frat{i\lambda}{N_c}\gamma_0(k,k)\right)+
\frat{i\lambda}{N_c}
\gamma_1(k,l)~\varphi(k-l)\right]~\psi(l)\right\}\nonumber
\end{eqnarray}
(we dropped out the factor normalizing to the free Lagrangian everywhere). 
It is pertinent to mention here the Diakonov-Petrov result comes to the play 
precisely if the scalar field is switched off.

In order to avoid a lot of the needless coefficients in the further 
formulae we introduce the  dimensionless variables 
(momenta, masses and vertices) 
\begin{equation}
\label{sc}
\frat{k\rho_c}{2}\to k~,~~~\frat{M\rho_c}{2}\to M~,
~~\gamma_0\to\rho_c^2\gamma_0~,~~~\gamma_1\to\rho_c\gamma_1~,
\end{equation}
the fields in turn 
\begin{equation}
\label{sc1}
\varphi(k)\to (n\kappa)^{-1/2}\rho_c^3\varphi(k)~,~~~~
\psi(k)\to \rho_c^{5/2}\psi(k)~,
\end{equation}
and eventually for $\lambda$ we are using 
$\mu=\frat{\lambda\rho_c^3}{2N_c}$. Then the generating functional 
takes the following form 
\begin{eqnarray}
\label{10}
&&{\cal Z}\simeq
\int d\mu~{\cal Z}^{''}_g\int D\psi^\dagger D\psi~D\varphi~
\exp\left\{-N\ln\mu-\int \frat{dk}{\pi^4}~\frat{1}{2}~\varphi(-k)~
4~[k^2+M^2]~\varphi(k)\right\}\times\nonumber\\
[-.2cm]
\\[-.25cm]
&&\times\exp\left\{\int \frat{dkdl}{\pi^8}~\psi^\dagger(k)
~2~\left[\pi^4\delta(k-l)(-\hat k+i\mu\gamma_0(k,k))+\frat{i\mu}
{(n\bar\rho^4\kappa)^{1/2}}
\gamma_1(k,l)~\varphi(k-l)\right]~\psi(l)\right\}~,\nonumber
\end{eqnarray}
where ${\cal Z}^{''}_g$ is a part of gluon component of the generating
functional which survives after expanding the action  per one instanton 
Eq. (\ref{dec}). The functional obtained describes the IL state influenced
by the quarks when all the terms containing the scalar field are collected
(see also Appendix). As mentioned above we believe this influence analogous 
to the reversal impact of phononlike deformations on the quark determinant
does not considerably change the numerical results of the IL and SCSB theory. 
But it is invisible from Eq. (\ref{10}) directly how this smallness is
reasoned. The scalar field enters the generating functional formally at 
the same order of the $\mu$-expansion as the term providing SCSB (compare
the second and third terms of the second exponential function in 
Eq. (\ref{10})). The suppression arises because the dominant contribution
of the Yukawa interaction comes from the quark field condensate which 
maintains
the additional $\mu$ smallness. This result prompts, in fact, rather
natural scheme of the approximate calculation of the generating functional
$$
\psi^\dagger\psi~\varphi=\langle\psi^\dagger\psi\rangle~\varphi+
\{\psi^\dagger\psi-\langle\psi^\dagger\psi\rangle\}~\varphi~.
$$
If we believe, for example, the first term ascribes some gluon component of 
the generating functional then being linearly dependent on the scalar field
it produces the small shift from the equilibrium value of the instanton 
size ($\rho_c\sim\bar\rho$) $\varphi=\varphi'+\delta\varphi$. And the shift
$\delta\varphi$ generates the mass term in the quark sector what means the
scheme should be pushed forward to the following expression
{\footnote{Let us emphasize the first term is responsible, in a sense, for 
the quark condensate fluctuations what allows the $\pi$ meson to gain its 
mass. In the second term such fluctuations are suppressed.}}
$$
\psi^\dagger\psi~\varphi=
\langle\psi^\dagger\psi\rangle~(\varphi'+\delta\varphi)+
\{\psi^\dagger\psi-\langle\psi^\dagger\psi\rangle\}~(\varphi'+\delta\varphi)~.
$$
Thus, we face the conventional mechanism of the mass generation 
when it is related to the insignificant variation of the equilibrium 
instanton size $\rho_c$ (or, in usual terms, to falling the scalar field 
condensate 
down) produced by the quark condensate. Moreover, it is clear if
the variations happen to be of the opposite sign then the field 
$\gamma_5\psi$
(the chiral partner of the field $\psi$) should develop the true mass. 
Before turning to the mechanism of the mass 
generation below let us calculate the quark
determinant integrating formally over the scalar field $D\varphi$,
then 
find the quark Green function in the tadpole approximation and formulate
the equation for the saddle point.
It is supposed the phonon component contribution does not affect 
substantially the results of the SCBS theory. Here we are interested in 
projecting upon the scale inherent for the scalar field.

\section*{III. Tadpole approximation}
The integration leads us to the four fermion interaction 
and the functional integral
can not be calculated exactly. However, due to 
smallness of scalar field corrections we may find the effective 
Lagrangian substituting the condensate value in lieu of one of the pairs of 
quark 
lines (see Fig. 1.)
\begin{figure}[htb]
\centerline{\epsfig{file=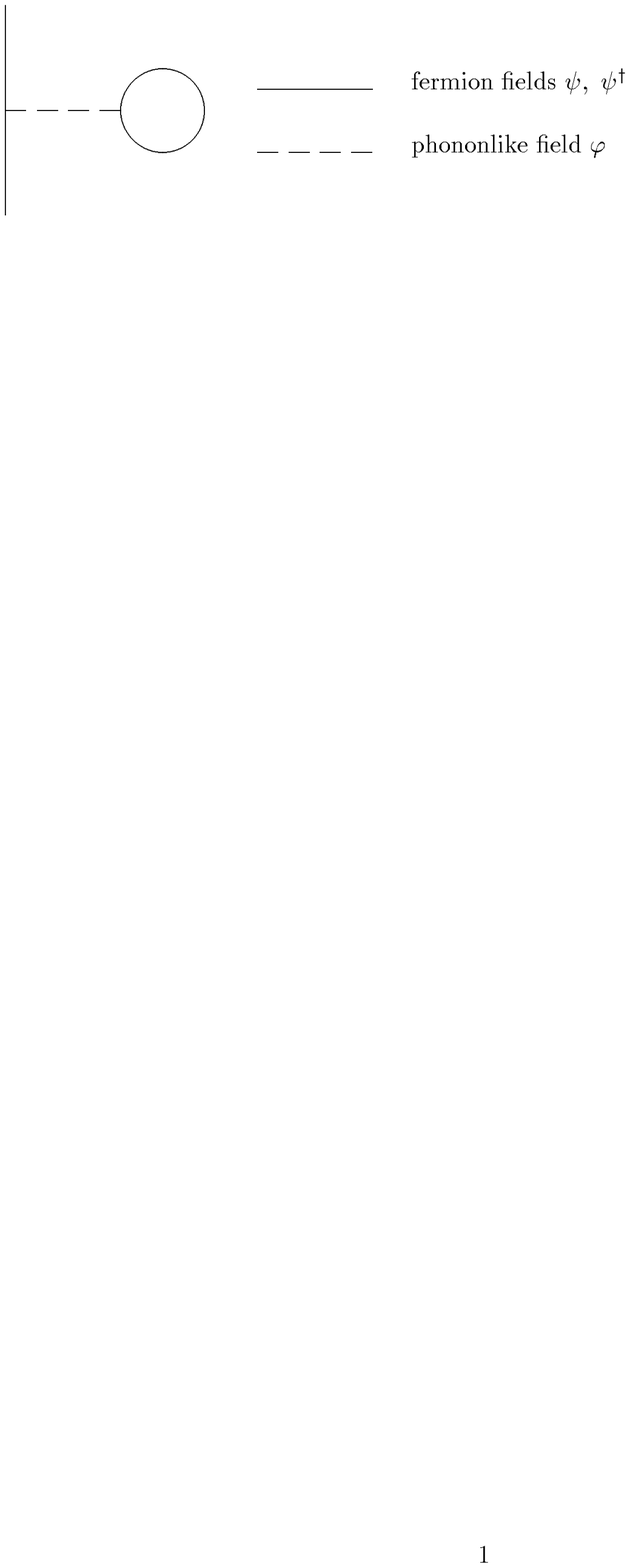,width=14cm}}
\vspace*{-16.5 cm}
\caption{The tadpole contribution.}
\end{figure}
$$\psi^\dagger (k)\psi(l)\to\langle\psi^\dagger (k)\psi(l)\rangle=
-\pi^4 \delta(k-l)~Tr~S(k)~.$$ 
In such an approach the diagram with four fermion lines in the lowest order 
of the perturbation theory in $\mu$ is reduced to the two-legs diagram
with one 
tadpole contribution (there are two such contributions because of two 
possible ways of pairing) 
\begin{eqnarray}
\label{12}
&&~~\frat{4~(i\mu)^2}{n\bar\rho^4\kappa}
\int \frat{dkdl~dk'dl'}{\pi^{16}}~\gamma_1(k,l)~\gamma_1(k',l')
~~\psi^\dagger(k)\psi(l)~~\psi^\dagger(k')\psi(l')~
~\langle \varphi(k-l)~\varphi(k'-l')\rangle \simeq\nonumber\\
&&\simeq\frat{4~\mu^2}{n\bar\rho^4\kappa}
\int \frat{dk}{\pi^4}~\gamma_1(k,k)~\psi^\dagger(k)\psi(k)
~\int \frat{dl}{\pi^4}~\gamma_1(l,l)~Tr~S(l)~D(0)~,\nonumber 
\end{eqnarray}
where the natural pairing definition was introduced
$$
\langle\varphi(k)~\varphi(l)\rangle=
\pi^4\delta(k-l)~D(k),~~D(k)=\frat{1}{4~(k^2+M^2)}~.
$$
It is obvious the factors surrounding $\psi^\dagger(k)\psi(k)$ has a 
meaning of quark mass
\begin{equation}
\label{kirmas}
m_f(k)=\frat{\mu}{(n\bar\rho^4\kappa)^{1/2}}~\gamma_1(k,k)
~\frat{(-2i\mu)}{(n\bar\rho^4\kappa)^{1/2}}
\int \frat{dl}{\pi^4}~\gamma_1(l,l)~Tr~S(l)~D(0)~,
\end{equation}
(the initial mass term contains the factor $2$ when the 
dimensionless variables are utilized, i.e. takes a form $2im_f$.) We are 
treating it as the current quark mass since the further calculations show
its magnitude gets just within the scale interval commonly accepted for the 
current quark mass and, moreover, it appears in the 
Gell-Mann--Oakes--Renner relation.

The contribution of the graph with all the quark lines paired (see Fig. 2) 
\begin{figure}[htb]
\centerline{\epsfig{file=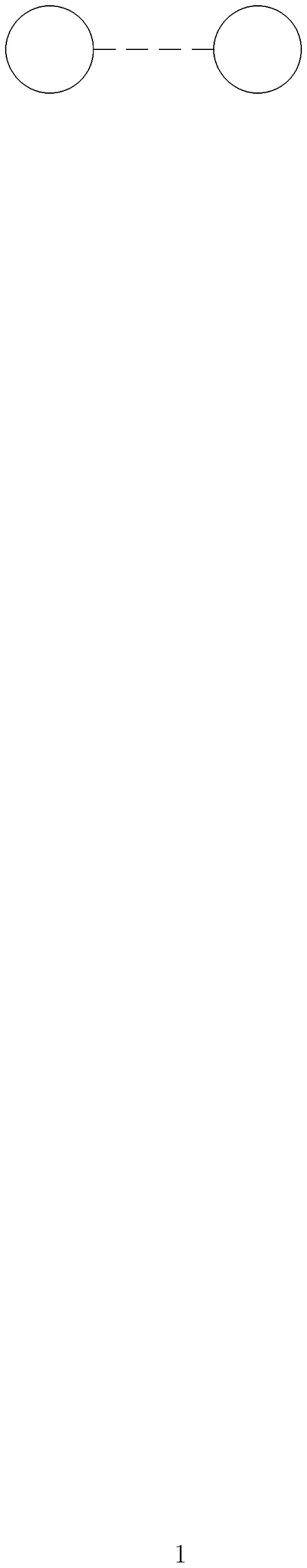,width=14cm}}
\vspace*{-16.5 cm}
\caption{Second order contribution.}
\end{figure}
should be taken into account at the same order of the $\mu$ expansion
while calculating the saddle point equation
$$
-\frat{2~\mu^2}{n\bar\rho^4\kappa}
\left[ \int \frat{dk}{\pi^4}~\gamma_1(k,k)~Tr~S(k)\right]^2
\pi^4\delta(0)D(0)=
-\frat12 \frat{\mu^2}{n\bar\rho^4\nu}~\frat{V}{\bar\rho^4}~
\left[ \int \frat{dk}{\pi^4}~\gamma_1(k,k)~Tr~S(k)\right]^{2}~.
$$
Here we used the natural regularization of the 
$\delta$-function $\delta(0)=\frat{1}{\pi^4}
\frat{V}{\bar\rho^4}$ in the dimensionless units. Then the quark
determinant after integrating over the scalar fields reads 
\begin{eqnarray}
\label{14}
&&{\cal Z}\sim \int d\mu\int D\psi^\dagger D\psi~
\exp\left\{-N\ln\mu+\frat{2N_c^{2}}{n\bar\rho^4\nu}~\frat{V}{\bar\rho^4}~
\mu^4~c^2(\mu)+\int\frat{dk}{\pi^4}~\psi^\dagger(k)~2~
[-\hat k+i\Gamma(k)]~\psi(k)\right\}=\nonumber\\
[-.2cm]
\\[-.25cm]
&&= \int d\mu~\exp\left\{-N\ln\mu+
\frat{2N_c^{2}}{n\bar\rho^4\nu}~\frat{V}{\bar\rho^4}~\mu^4~c^2(\mu)
+~\frat{V}{\bar\rho^4}~\int\frat{dk}{\pi^4}~Tr~
\ln[-\hat k+i\Gamma(k)]\right\}~,\nonumber
\end{eqnarray}
where the vertex function is defined as
$$
\Gamma(k)=\mu~\gamma_0(k,k)+m_f(k)~,
$$
and we introduced the function $c(\mu)$ convenient for the practical
calculations 
$$
c(\mu)=-\frat{i}{2\mu~N_c}~ \int\frat{dk}{\pi^4}~\gamma_1(k,k)~Tr~S(k)~.
$$
\begin{figure}[htb]
\grpicture{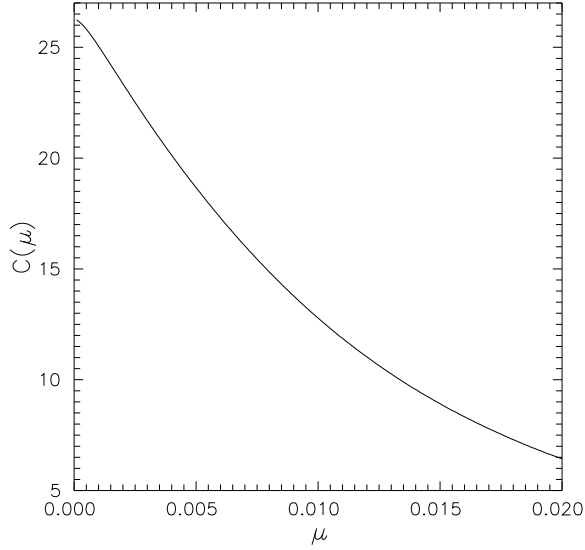}
%\centerline{\epsfig{file=cmu.ps,width=5cm}}
%\vspace*{.5 cm}
\caption{The function $c(\mu)$ at $N_f=1$.}
\end{figure}  
As it is clear from Eq. (\ref{14}) the Green function of the quark field is
self-consistently defined by the following equation  
$$
2~[-\hat k+i\Gamma(k)]~S(k)=-1~.
$$
Searching the solution in the form
$$
S(k)=~A(k)~\hat k+i~B(k)~,
$$
we get
$$A(k)=\frat12~\frat{1}{k^2+\Gamma^2(k)}~,~~~~~~
B(k)=\frat12~\frat{\Gamma(k)}{k^2+\Gamma^2(k)}~.
$$
Using Eq. (\ref{kirmas}) and the definitions of $\Gamma(k)$ and $B(k)$ 
we have the complete integral equation
$$
\Gamma(k)=\mu~\gamma_0(k,k)+\frat{N_c}{n\bar\rho^4\nu}~\mu^2~\gamma_1(k,k)
~\int\frat{dl}{\pi^4}~\gamma_1(l,l)~\frat{\Gamma(l)}{l^2+\Gamma^2(l)}~,
$$
which drives to have the convenient representation of the solution
$$
\Gamma(k)=\mu~\gamma_0(k,k)+\frat{N_c}{n\bar\rho^4\nu}~\mu^3~c(\mu)
~\gamma_1(k,k)~.
$$ 
What concerns the function $c(\mu)$ it is not a great deal to obtain 
$$
c(\mu)=\frat{1}{\mu}
~\int\frat{dk}{\pi^4}~\gamma_1(k,k)~\frat{\Gamma(k)}{k^2+\Gamma^2(k)}~,
$$ 
and, therefore, the complete integral equation for the function
{\footnote{
The solution is unique within the $\mu$ interval of our interest. It is 
curious to notice when $\mu$ is larger than $\sim \dmn{4}{-2}$ several 
solution branches for $c(\mu)$  emerge (see Fig. 4). 
Surely, it could be interesting
to clarify if there is any correspondence between these branches and the
saddle point equation resulting from Eq. (\ref{14}). However, it is clear  
wittingly those objects should be heavier than the scale of several
hundred $MeV$ characteristic for SCSB. It is quite possible such
solutions might be associated with some 'heavy' particles if they do 
exist.}}
$c(\mu)$ which is shown in Fig. 3 for $N_f=1$.
Let us underline the $N_f$-dependence of the $c(\mu)$ function in the 
interval of $\mu$ determined by saddle point value is unessential. 
Then we easily obtain for the current quark mass
\begin{figure}[htb]
\grpicture{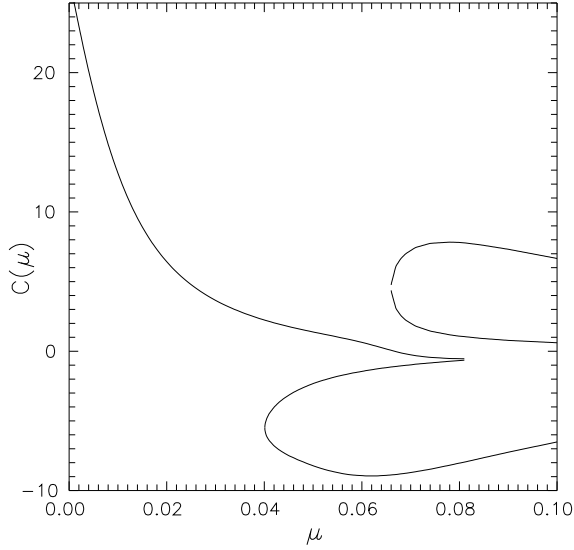}
%\centerline{\epsfig{file=Root.ps,width=14cm}}
%\vspace*{-6.5 cm}
\caption{The branches of the function $c(\mu)$.}
\end{figure}   
\begin{equation}
\label{mff}
m_f(k)=\frat{N_c}{n\bar\rho^4}~\frat{\mu^3}{\nu}~c(\mu)~\gamma_1(k,k)~,
\end{equation}
and see the cancellation of the kinetic coefficient $\kappa$ 
in $m_f$. Thus, it means the precise value of the coefficient is 
unessential as declared. 

We have the following equation for the saddle point of the 
functional of Eq. (\ref{14}) 
\begin{equation}
\label{20}
\frat{2N_c}{n\bar\rho^4}
\int\frat{dk}{\pi^4}~\frat{\mu~[\Gamma^2(k)]'_{\mu}}{k^2+\Gamma^2(k)}+
\frat{2N_c^{2}}{n\bar\rho^4\nu}~\frat{\mu}{n\bar\rho^4}~
[\mu^4~c^2(\mu)]'_{\mu}=1~,
\end{equation} 
where the prime is attributed to the differentiation in $\mu$.

It results from assuming the stationary IL parameters, 
what is not accurate. We should include another effect 
produced by the shift of equilibrium instanton size thanks to the quark 
condensate presence. The modification of the IL parameters ($n(\mu),\cdots$)
caused by the Yukawa interaction comes from simple tadpole graph of Fig. 5 
\begin{figure}[htb]
\centerline{\epsfig{file=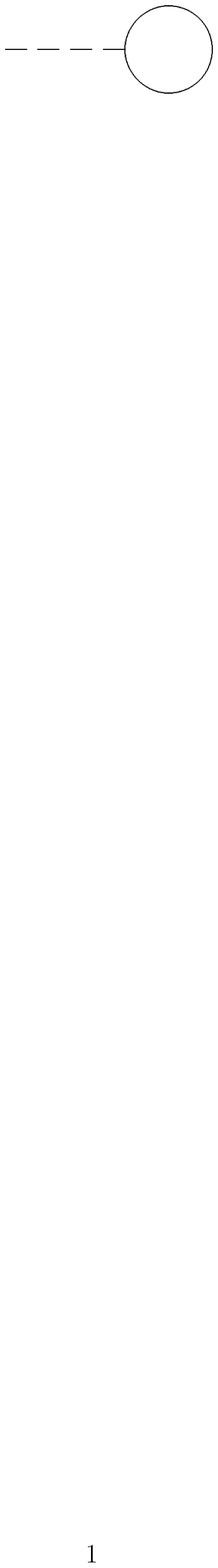,width=14cm}}
\vspace*{-16.5 cm}
\caption{The tadpole.}
\end{figure}
in the leading order 
\begin{eqnarray}
\label{tdp}
&&\frat{2~i\mu}{(n\bar\rho^4\kappa)^{1/2}}
\int \frat{dkdl}{\pi^8}~\gamma_1(k,l)~(-\pi^4)~\delta(k-l)~Tr~S(k)
~\varphi(k-l)=\Delta\cdot~\varphi(0)~,\nonumber\\
[-.2cm]
\\[-.25cm]
&&\Delta=-\frat{2i\mu}{(n\bar\rho^4\kappa)^{1/2}}
\int \frat{dk}{\pi^4}~\gamma_1(k,k)~Tr~S(k)=
\frat{4N_c}{(n\bar\rho^4\kappa)^{1/2}}~\mu^2 c(\mu)~,\nonumber
\end{eqnarray}
and this term generates the small shift of the equilibrium 
instanton size $\rho_c\sim\bar\rho$ (remind here $\varphi=\rho-\rho_c$, 
and $\varphi(0)=\int dz~\varphi(z)$ is the scalar field in momentum
representation)
{\footnote{This shift in phononlike component of Lagrangian
$-4~\frat{M^2}{2}\varphi^2+\Delta\cdot~\varphi=-4~\frat{M^2}{2}
\varphi'^2+\cdots$
with the definitions $\varphi'=\varphi-\delta\varphi$ and
$\delta\varphi=\frat{\Delta}{4~M^2}$
generates the mass term in quark sector
$m^{'}_f(k)=\frat{\mu}{(n\bar\rho^4\kappa)^{1/2}}~
\gamma_1(k,k)~\delta\varphi$,
which has obviously the same form as that of Eq. (\ref{mff}).}}.   
The tadpole contribution could be absorbed in making more sophisticated  
variational procedure of saddle point calculation. 
Actually, it includes also the variation of the IL parameters as a 
function of $\mu$. But in practice it comes about highly effective  to 
use the simple iterating procedure. At the first step this variation of the 
IL parameters is 
not taken into account and the saddle point $\mu(\rho_c)$ is calculated 
from Eq. (\ref{20}). Then getting new IL parameters (see, Appendix) we have 
to resolve Eq. (\ref{20}) again and etc. Five or six iterations are quite 
enough if we are satisfied with the same precision as in calculating the 
integrals.  
In the Table 1 the numerical results  (M.S.Z.) are shown for 
$N_f=0,1$ ($N_f=0$ corresponds to the quenched 
approximation for the 
IL parameters) comparing to those of Diakonov and Petrov (D.P.) 
where the disturbance of instanton medium was not considered. 
\begin{center}
Table 1.\\\vspace{0.3cm}
\begin{tabular}{|cccc|ccccc|}
\hline
     &     &D.P.  &                                  &         
 &      &M.S.Z.&                                        &\\\hline
$N_f$&$\mu$&$M(0)$&$-i\langle\psi^\dagger\psi\rangle$
 &$N_f$ &$\mu$ &$M(0)$&$-i\langle\psi^\dagger\psi\rangle$&
$m_f$\\\hline
$0$ &\dmn{5.68}{-3}&$341$&$-(301)^3$&
$0$ &\dmn{5.51}{-3}&$366$&$-(297)^3$&$4.29$\\      
$1$ &\dmn{5.14}{-3}&$376$&$-(356)^3$&
$1$ &\dmn{5.07}{-3}&$385$&$-(328)^3$&$5.13$\\
\hline
\end{tabular}
\end{center}
The parameters indicated in the Table 1 are\\ 
the dynamical quark mass\\
$$
 M(0)= \Gamma(0)~\left(\frat{2}{\rho_c}\right)~~~[MeV]~,
$$
the quark condensate\\
$$-i\langle\psi^\dagger\psi\rangle=-i~Tr~S(x)|_{x=0}=\frat{N_c}{4}~
\int \frat{dk}{\pi^4} \frat{\Gamma(k)}{k^2+\Gamma^2(k)}
~\left(\frat{2}{\rho_c}\right)^3~~~[MeV]^3~,
$$
and, at last, the current quark mass $m_f~~[MeV]$, which is defined now
being mixed with the quark condensate as
\begin{equation}
\label{mcur}
m_f=\frat{\int dk~m_f(k)~Tr~S(k)}
{\int dk~Tr~S(k)}~.
\end{equation}
It is just what is dictated by the Gell-Mann--Oakes--Renner relation which 
we calculate below. 
Through this paper the value of renormalization 
constant is fixed by $\Lambda=280~MeV$. 
Then the IL parameters are slightly different from their 
conventional values 
$\bar\rho\sim (600~MeV)^{-1},~\bar R\sim (200~MeV)^{-1}$ 
(see the corresponding tables in Appendix). 
However, with the minor $\Lambda$ variation the parameters could be
optimally fitted. 
As expected the change of quark condensate is insignificant, the
order of several $MeV$, what hints the existence of new soft energy scale  
established by the disturbance which accompanies the quark propagation through
the instanton medium.

\section*{IV. Multiflavour approach}
In order to match the approach developed to phenomenological estimates
we need the generalization for $N_f>1$.
Then the quark determinant becomes
\cite{2}, \cite{dp2} 
$$
{\cal Z}_\psi~\simeq~\int D\psi^\dagger D\psi~ 
\exp\left\{\int d x~\sum_{f=1}^{N_f}\psi_{f}^\dagger(x)~
i\hat\partial\psi_{f}(x)\right\}~
\left(\frat{Y^+}{VM^{N_f}}\right)^{N_+}~
\left(\frat{Y^-}{VM^{N_f}}\right)^{N_-}~,
$$
$$Y^{\pm}=i^{N_f} \int dz~dU~d\rho~n(\rho)/n
\prod_{f=1}^{N_f}\int dx_fdy_f~\psi_{f}^\dagger(x_f)~i\hat\partial_{x_f} 
\Phi_\pm(x_f-z)~
\Phi^\dagger_{\pm}(y_f-z)~i\hat\partial_{y_f}~\psi_{f}(y_f)~. 
$$
With phononlike component included every pair of the zero modes 
$\sim\Phi~\Phi^\dagger$ acquires the additional term similar to 
Eq. (\ref{8}). The appropriate transformation driving the factors $Y^\pm$ 
to their determinant forms \cite{2} is still valid here since the 
correction term differs from the basic one with the  scalar field 
$\varphi$.  
The complete integration over $dz$ leads (in the adiabatic approximation 
$\varphi(x,z)\to\varphi(z)$) to the transparent Lagrangian form with the 
momentum conservation of all interacting particles. 
Besides, we keep the main terms of $Y^\pm$ in $\varphi$ expansion.
The quark zero modes generate the factor similar to Eq. (\ref{8a}) with  
$\frat{1}{N_c}$ being changed by the factor 
$\left(\frat{1}{N_c}\right)^{N_f}$ and then in the leading  $N_c$ order
we have 
\begin{eqnarray}
&&Y^\pm=\left(\frat{1}{N_c}\right)^{N_f}\int dz~
\det_{N_f}\left(i~J^{\pm}(z)\right)~,\nonumber\\
&&J^{\pm}_{fg}(z)=\int\frat{dkdl}{(2\pi)^8}
\left[e^{i(k-l)z}\gamma_0(k,l)+\int\frat{dp}{(2\pi)^4}~e^{i(k-l+p)z}
\gamma_1(k,l)~\varphi(p)\right]~
\psi^\dagger_{f}(k)~\frat{1\pm\gamma_5}{2}~\psi_g(l)~.
\nonumber
\end{eqnarray}
While providing the Gaussian form for the functional we perform
the integration over the auxiliary parameter $\lambda$ together with the 
bosonization resulting in the integration over the auxiliary matrix 
$N_f\times N_f$ meson fields \cite{dp2}
$$
\exp\left[\lambda~\det \left(\frat{i~J}{N_c}\right)\right]\simeq
\int d {\cal M}~\exp\left\{i~Tr[{\cal M}J]-(N_f-1)
\left(\frat{\det[{\cal M}N_c]}{\lambda}\right)^{\frac{1}{N_f-1}}
\right\}~.
$$
As a result the generating functional may be written as
\begin{eqnarray}
\label{24}
&&{\cal Z}=\int\frat{d\lambda}{2\pi}~{\cal Z}^{''}_g~
\exp(-N\ln\lambda)~\int D\varphi~
\exp\left\{-\int \frat{dk}{(2\pi)^4}~\frat{n\kappa}{2}~\varphi(-k)~
[k^2+M^2]~\varphi(k)\right\}\cdot\nonumber\\
&&\cdot \int D{\cal M}_{L,R}~\exp\int dz
\left\{-(N_f-1)\left[
\left(\frat{\det[{\cal M}_L N_c]}{\lambda}\right)^{\frac{1}{N_f-1}}+
\left(\frat{\det[{\cal M}_R N_c]}{\lambda}\right)^{\frac{1}{N_f-1}}
\right]\right\}\cdot\\
&&\cdot\int D\psi^\dagger D\psi~\exp 
\left\{\int \frat{dk}{(2\pi)^4} \sum_f\psi_{f}^\dagger(k)(-\hat k)\psi_f(k) 
+i~\int dz\left(Tr[{\cal M}_L J^+]+Tr[{\cal M}_R J^-]\right)\right\}~.
\nonumber
\end{eqnarray}
Now  scalar field interacts with the quarks of the different 
flavours, nevertheless,  the dominant contribution is expected  from 
the tadpole graphs where any pair of the quark fields is taken in the 
condensate approximation as it happens at $N_f=1$
$$
\psi^\dagger_{f}(k)\psi_g(l)\to\langle\psi^\dagger_{f}(k)\psi_g(l)\rangle=
-\pi^4\delta_{fg}~\delta(k-l)~Tr~S(k)~.
$$
As for the condensate itself we obtain it as the nontrivial solution
of saddle point equation. For example, it is
$$({\cal M}_{L,R})_{fg}={\cal M}~\delta_{fg}
$$
for the diagonal meson fields.
The dimensionless convenient variables (in addition to  
Eqs. (\ref{sc}), (\ref{sc1})) are the following
$$
\frat{{\cal M}}{2}~\bar\rho^3\to \mu~,
~~~~\left(\frat{\lambda\bar\rho^4}{(2N_c\bar\rho)^{N_f}}
\right)^{\frac{1}{N_f-1}}
\to g~.
$$
Then the effective action (${\cal Z}=\int dgd\mu~\exp \{-V_{eff}\}$) in new
designations has the form
$$
V_{eff}=N(N_f-1)\ln g-\frat{V}{\bar\rho^4}(N_f-1)
\frat{2\mu^{\frac{N_f}{N_f-1}}}{g}
-\frat{V}{\bar\rho^4}\frat{2N_f^{2}N_c^{2}}{n\bar\rho^4~\nu}\mu^4 c^2(\mu)-
2N_fN_c~\frat{V}{\bar\rho^4}~\int \frat{dk}{\pi^4}\ln\{k^2+\Gamma^2(k)\},
$$
and  the saddle point equation reads 
$$
\frat{2N_c}{n\bar\rho^4}
\int\frat{dk}{\pi^4}~\frat{\mu~[\Gamma^2(k)]'_{\mu}}{k^2+\Gamma^2(k)}+
\frat{2~N_f~N_c^{2}}{n\bar\rho^4\nu}~\frat{\mu}{n\bar\rho^4}~
[\mu^4~c^2(\mu)]'_{\mu}=1~.
$$
The quark current mass in Eq. (\ref{mff})
gains the additional factor $N_f$
because the scalar nature of the phononlike
field  requires to match the tadpole quark field condensates of all
$N_f$ flavours to every vertex
$$
m_f(k)=\frat{N_f N_c}{n\bar\rho^4}~\frat{\mu^3}{\nu}~c(\mu)~\gamma_1(k,k)~.
$$
Table 2 complements the Table 1 with the calculations at $N_f=2$
\begin{center}
Table 2.\\\vspace{0.3cm}
\begin{tabular}{|ccccc|cccccc|}
\hline
&D.P.&&&&&M.S.Z.&&&&\\\hline
$\mu$&$M(0)$&$-i\langle\psi^\dagger\psi\rangle$&$f_{\pi}$&$f^{'}_\pi$
&$\mu$&$M(0)$&$-i\langle\psi^\dagger\psi\rangle$&$f_{\pi}$&$f^{'}_\pi$&
$m_f$\\\hline
\dmn{4.66}{-3}&$422$&$-(427)^3$&$135$&$111$
&\dmn{4.61}{-3}&$416$&$-(374)^3$&$118$&$97.2$&$12.7$
\\\hline
\end{tabular}
\end{center}
where $f_{\pi}~~[MeV]$ is the pion decay constant and
$$
f_{\pi}^2=~\frat{N_c N_f}{8}~
\int \frat{dk}{\pi^4}~\frat{\Gamma^2(k)-\frat{k}{2}\Gamma'(k)\Gamma(k)
+\frat{k^2}{4}(\Gamma'(k))^2}{(k^2+\Gamma^2(k))^2}
~\left(\frat{2}{\rho_c}\right)^2~,
$$
$f^{'}_\pi~~[MeV]$ is its approximated form 
$$
f^{'2}_\pi=\frat{N_c N_f}{8}~
\int \frat{dk}{\pi^4} \frat{\Gamma^2(k)}{(k^2+\Gamma^2(k))^2}
~\left(\frat{2}{\rho_c}\right)^2~,
$$
here $\Gamma'(k)=\frat{d\Gamma(k)}{dk}$, 
and the condensate $-i\langle\psi^\dagger \psi\rangle$ is implied for the 
quarks of every flavour.

\section*{V. Meson excitations of the chiral condensate}
Here we discuss the meson excitations of the chiral condensate
adapting the effective Lagrangian to the suitable variables. 
The meson degrees of freedom are introduced into the effective
Lagrangian (\ref{24}) with $N_f=2$ by the following substitutions 
\begin{eqnarray}
\label{29}
&&{\cal M}^L={\cal M}~\widetilde{\cal M}^L~,~
\widetilde{\cal M}^L=(1+\sigma+\eta)~U~V~,
~~~~U=\exp(i\pi_a\tau^a)~,~~~~~V=\exp(i\sigma_a\tau^a),\nonumber\\
&&{\cal M}_R={\cal M}~\widetilde{\cal M}^R~,
~\widetilde{\cal M}^R=(1+\sigma-\eta)~V~U^\dagger~,\nonumber
\end{eqnarray}
where ${\cal M}$ denotes the condensate and $\sigma,~\eta$ are the scalar
and pseudoscalar meson fields, $\pi^a$ is the isotriplet of the $\pi$-mesons,
$\sigma^a$ is the vector meson isotriplet. The effective action for the meson 
field excitations looks then as
$$
V_{eff}=-\int\frat{dp}{\pi^4}~\{
\pi^a(p)~R_{\pi^a}(p)~\pi^a(-p)+
\sigma^a(p)~R_{\sigma^a}(p)~\sigma^a(-p)+
\sigma(p)~R_\sigma(p)~\sigma(-p)+
\eta(p)~R_\eta(p)~\eta(-p)
\}~,
$$
here $R_{\pi^a},~R_{\sigma^a},~R_\sigma,~R_\eta$ are the inverse propagators
of the corresponding particles. Their exact forms are 
listed in Ref. \cite{2} but for the $\pi$- and 
$\sigma$-mesons with the phononlike contributions included they are
calculated below.

The contributions generated by the quark determinant if the phononlike
fields ignored are well investigated in the leading order of the 
$N_c$-expansion. 
In particular, one of the contributions  comes from the diagram A).
\begin{figure}[htb]
\centerline{\epsfig{file=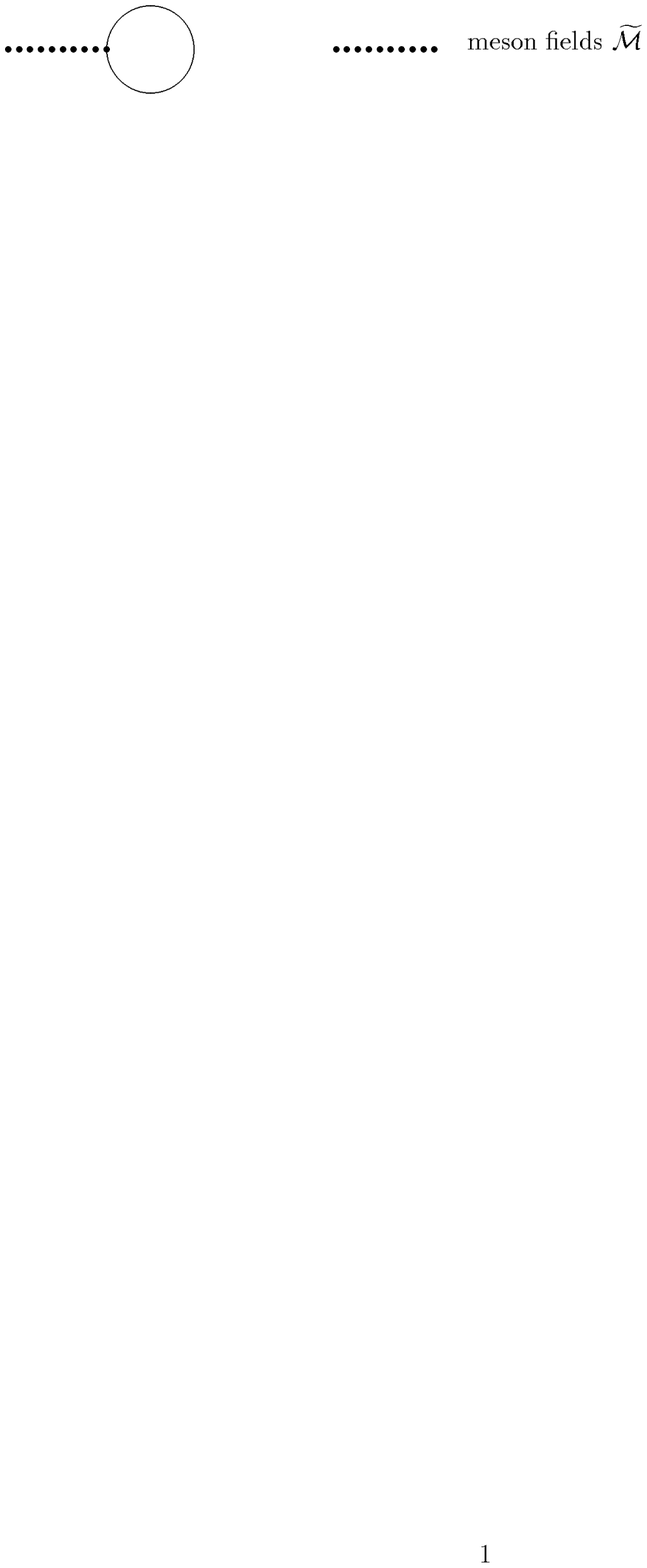,width=14cm}}
\vspace*{-16.5 cm}
\caption{Diagram A).}
\end{figure}
It is  of the first order in the 
$\mu$-expansion with one quark loop and one external meson leg in which  
the meson fields are ~maintained  up to the quadratic terms 
\begin{eqnarray}
\label{A}
A)~~~2i~\int dz\int\frat{dk}{\pi^4}~(-Tr~S(k))~\Gamma(k)~\left\{
\widetilde {\cal M}^L_{ff}(z)~\frat{1+\gamma_5}{2}+
\widetilde {\cal M}^R_{ff}(z)~\frat{1-\gamma_5}{2}\right\}~.\nonumber
\end{eqnarray}
Besides, another contribution comes from the diagram B) 
being of the second order in $\mu$ with one quark loop and two external
meson legs
\begin{figure}[htb]
\centerline{\epsfig{file=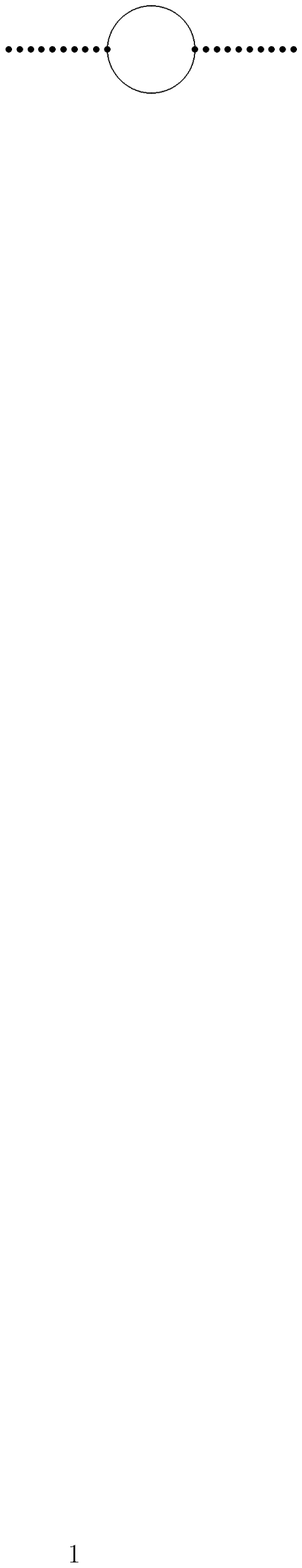,width=14cm}}
\vspace*{-16.5 cm}
\caption{Diagram B).}
\end{figure}
\begin{eqnarray}
\label{B}
B)~ 
2\int \frat{dkdl}{\pi^8}Tr[S(k)S(l)]~\Gamma(k,l)\Gamma(l,k)~
\left\{\widetilde {\cal M}^L_{fg}(l-k)\frat{1+\gamma_5}{2}
\widetilde {\cal M}^L_{gf}(k-l)\frat{1+\gamma_5}{2}+
(L,\gamma_5)\to (R,-\gamma_5)\right\},\nonumber
\end{eqnarray} 
where we introduced $\Gamma(k,l)=\mu\gamma_0(k,l)+
\frat{\mu}{(n\bar\rho^4\kappa)^{1/2}}~\gamma_1(k,l)~\delta\varphi,~
\gamma_0(k,l)=G(k)G(l)$.
The determinant terms of the generating functional at $N_f=2$ develop 
the form
$$
det~\widetilde {\cal M}^L+det~\widetilde {\cal M}^R=1+\sigma^2+\eta^2~.
$$
In addition to this diagrams we have the tadpole diagram C)
in which the meson fields are maintaned up to the quadratic 
terms and which describes the effect of equilibrium instanton size 
fluctuations when the meson fields are present. 
\begin{figure}[htb]
\centerline{\epsfig{file=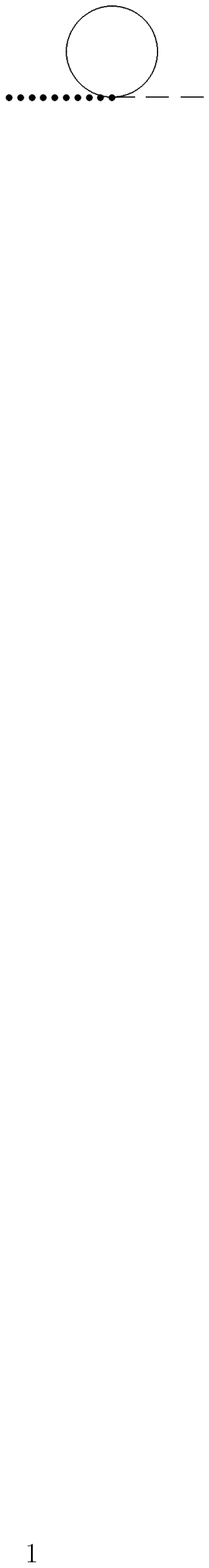,width=14cm}}
\vspace*{-16.5 cm}
\caption{Diagram C).}
\end{figure}
These fluctuations influence  the IL parameters
{\footnote{
The problems related to the effective chiral Lagrangian (when the phononlike
fields included) and its symmetries will be discussed in the separate 
publication.
}}.
\begin{eqnarray}
\label{D}
C)~~~\int \frat{dk}{\pi^4}~Tr~(-S(k))~2i~m_f(k)~\int dz
\left\{\widetilde {\cal M}^L_{ff}(z)~\frat{1+\gamma_5}{2}+
\widetilde {\cal M}^R_{ff}(z)~\frat{1-\gamma_5}{2}\right\}~.\nonumber
\end{eqnarray}
In particular, we have for the inverse propagator of the $\pi$-meson field
\begin{eqnarray}
\label{pim}
&&R_{\pi^a}(p)=\int\frat{dk}{\pi^4}~\Big\{ -2~N_c~N_f~\frat{(k,k+p)+
\Gamma(k)\Gamma(k+p)}
{[k^2+\Gamma^2(k)][(k+p)^2+\Gamma^2(k+p)]}~\Gamma(k,k+p)~\Gamma(k+p,k)+
\nonumber\\
&&+2 N_c~N_f~\frat{\Gamma^2(k)}{k^2+\Gamma^2(k)}
+i~N_f~m_f(k)~Tr~(-S(k))~\Big\}~.\nonumber
\end{eqnarray}
When the pion momentum goes to zero $p\to 0$ the first two
terms do not generate the massive term because of
the quark determinant symmetry. As a result, these terms at small values 
of $p$ originate only term proportional to $p^2$
and we limit ourselves
with the approximate result for $\Gamma=\mu\gamma_0$ while calculating it.
Using the expansion obtained in Ref. \cite{2} we have
\begin{equation}
\label{rez}
R_{\pi^a}(p)=i~N_f~\int \frat{dk}{\pi^4}~
m_f(k)~Tr~(-S(k))+\beta n~p^2
\end{equation}
with
$\beta n=\frat{N_c~N_f}{16} 
\int \frat{dk}{\pi^4}~\frat{\Gamma^2(k)-\frat{k}{2}\Gamma'(k)\Gamma(k)
+\frat{k^2}{4}(\Gamma'(k))^2}{(k^2+\Gamma^2(k))^2}~
\left(\frat{2}{\rho_c}\right)^2$.
Eq. (\ref{rez}) combined with the definition (\ref{mcur}) can be given as 
$$R_{\pi^a}(p)=~\beta n~\left\{~\frat{m_f~i\langle \psi^\dagger
\psi\rangle N_f}{\beta n}+p^2\right\}~.
$$
From here we have for the $\pi$-meson mass
$m^{2}_\pi=\frat{m_f~i\langle \psi^\dagger\psi\rangle N_f}
{\beta n}$
or if the relation between $\beta$ and the pion constant $f_\pi$ 
($f_\pi^{2}=2\beta n$) plugged in we obtain
\begin{equation}
\label{masfi}
m^{2}_\pi=\frat{2~m_f~i\langle \psi^\dagger\psi\rangle N_f}
{f^{2}_\pi}~,
\end{equation}
what displays Gell-Mann--Oakes--Renner relation
$$m^{2}_\pi=\frat{(m_u+m_d)~i[\langle u^\dagger u\rangle +
\langle d^\dagger d\rangle]}
{f^{2}_\pi}~.$$
The factor $2$ in the numerator of Eq. (\ref{masfi}) corresponds just the 
sum of the $u$ and $d$ quark masses and $N_f$ means the summation of 
condensates. 

The light particle which we introduced and which imitates the scalar glueball
properties does not affect significantly the SCSB parameters and correctly
describes the soft pion excitations of quark condensate. Meanwhile, the 
experimental status of this light scalar glueball is very vague. We 
believe the phononlike excitations could manifest themselves being mixed
with the excitations of the quark condensate in the scalar channel. To 
illuminate the point let us consider the inverse propagator of the 
$\sigma$-meson which is given by the contributions of the diagram $B)$ and 
determinant
\begin{eqnarray}
\label{sig}
R_\sigma(p)=-2~N_c~N_f~\int\frat{dk}{\pi^4}~ 
\frat{(k,k+p)-\Gamma(k)\Gamma(k+p)}
{[k^2+\Gamma^2(k)][(k+p)^2+\Gamma^2(k+p)]}~\Gamma(k,k+p)~\Gamma(k+p,k) +
n\bar\rho^4~.\nonumber
\end{eqnarray}
Holding the highest terms of the $\mu$-expansion only, when 
$\Gamma=\mu\gamma_0$, we receive the following result by means of the 
identity (see, \cite{2}) 
\begin{eqnarray}
\label{sigm}
R_\sigma(p)=N_c~N_f~\int dk~ 
\frat{[~\Gamma(k)~(k+p)_\mu+\Gamma(k+p)~k_\mu~]^2}
{[k^2+\Gamma^2(k)][(k+p)^2+\Gamma^2(k+p)]}
~.\nonumber
\end{eqnarray}
In the course of this exercise we have to include the contribution of the
'shifting' diagram of  $C)$ type where the field $\varphi$ (more exactly 
$\varphi'$) should be treated as the dynamical one, i.e.
\begin{eqnarray}
\label{D'}
C')~~~2i\mu\int \frat{dk}{\pi^4}~Tr~(-S(k))~
\frat{\gamma_1(k,l)}{(n\bar\rho^4)^{1/2}}
\int dz~\varphi(z)\left\{\widetilde {\cal M}^L_{ff}(z)\frat{1+\gamma_5}{2}+
\widetilde {\cal M}^R_{ff}(z)\frat{1-\gamma_5}{2}\right\}.\nonumber
\end{eqnarray}
Then the term of interacting scalar fields which we are interested in looks
like
$V_{\varphi\sigma}=
\int dz~ \Delta~ \varphi(z) ~\sigma(z)$,
where $\Delta$ is defined by Eq. (\ref{tdp}).
At low momenta the diagrams contribute to the effective action 
(in the dimensionless variables) as
$$
-2~(p^2+M^2)~\varphi^2-
2~n\bar\beta_\sigma~(p^2+M^2_{\sigma})~\sigma^2+
\Delta~ \varphi~\sigma
$$
with $\bar\beta_\sigma$ standing for the kinetic coefficient of the 
$\sigma$-meson and $M_{\sigma}$ is its mass. Both quantities are derived
from the expansion of the inverse propagator in the low momentum region as
$R_\sigma(p)\simeq 2~n\bar\beta_\sigma~(p^2+M^2_{\sigma})$.
Diagonalizing the quadratic form
$$
-2~(p^2+M^2)~\varphi^2-2~(p^2+M^2_{\sigma})~\widetilde\sigma^2+
\widetilde\Delta ~\varphi~\widetilde\sigma~,
$$
(presented in more adequate variables
$\widetilde\sigma=(n\bar\beta_\sigma)^{1/2}~\sigma,~
\widetilde\Delta=\frat{\Delta}{(n\bar\beta_\sigma)^{1/2}}$)
leads to the following definition of the composite particle masses
\begin{equation}
\label{simas}
M^2_{1,2}=\frat{M^2_{\sigma}+M^2}{2}\pm
\frat{\sqrt{[M^2_{\sigma}-M^2]^2+\widetilde\Delta^2}}{2}~.
\end{equation}
It is well known that SCSB is inapplicable when we are going to deal with
heavy meson masses. Specifically, it leads to the wrong predictions for the
$\sigma$-meson because the corresponding mass obtained is of an order 
$1/\bar\rho\sim 1~GeV$ (moreover, the adiabaticity approximation is certainly
broken then). The result Eq. (\ref{simas}) aggravates, in a sense,
the situation since the hard component leaves for the region of  harder 
masses whereas the light component persists to become lighter. On the other
hand, the present observations signal rather the existence of the scalar 
meson of mass about
$0.5~GeV$  what apparently does not coincide with the light
component $M_2$.
Nevertheless, our consideration having no claims of the quantitative agreement 
shows the scalar meson could be the mixed particle of pretty large width
(similar to the superposition of oscillators with the various fundamental
frequencies) describing the excitations of chiral and
gluon condensates.

\section*{VI. Conclusion}
In this paper we have developed the consistent approach to describe the 
interaction of quarks with IL. Theoretically, it is based (and justified) 
on the particular choice of the configurations saturating the functional 
integral what occurs to be not merely a technical exercise. They are the 
deformable (crampled) (anti-)instantons with the variable parameters 
$\gamma(x)$ and in the concrete treatment of this paper we play with the 
variation of the PP size $\rho(x,z)$. In a sense, such an ansatz is strongly 
motivated by the form of quark determinant which is solely dependent on the 
average instanton size in the SCSB theory. We have demonstrated  that in the 
long-length wave approximation the variational problem of the deformation 
field optimization turns into the construction of effective Lagrangian 
for the  scalar phononlike  $\varphi$ and quark fields with the Yukawa 
interaction. Physically, it allows us to analyze the inverse influence of 
quarks on the instanton vacuum. We have pointed out this influence on the IL 
parameters as negligible. The modification of the SCSB parameters occurs 
pretty poor as well. In particular, the scale of quark condensate change 
amounts to a few $MeV$ only. Nevertheless, switching on the phononlike 
excitations of IL leads to several qualitatively new and interesting effects. 
The propagation of the quark condensate disturbances over IL happens in this 
approach to be in close analogy
with well-known polaron problem. We imply a necessity to take into account
the medium feedback while elementary excitations  propagating.
The intriguing conclusion  comes with realizing it leads to generation
of current quark masses $m_f$ within QCD itself with their values
entirely corresponding to the conventional phenomenological results for $u$
and $d$ quarks. Moreover, the $\pi$-meson being massless pseudo-Goldstone 
particle in the standard SCSB theory acquires its mass obeying the 
Gell-Mann--Oakes--Renner relation when the IL deformations enter the game. 
Besides, it hints that fitting the parameters $\bar\rho,~n$ and 
renormalization constant $\Lambda$ all together with the alteration of 
$s(\rho)$ profile function we might achieve suitable agreement not only in 
the order of magnitude. The difficulties which confronted us here illuminate 
the fundamental problem of gluon field penetration into the vacuum 
(the instanton vacuum in this particular case) as the most principle one. 
Indeed, it is a real challenge to answer the  question about the strong 
interaction carrier in the soft momentum region. Perhaps, the light particle 
of scalar glueball properties which appears inherently in our approach and 
should manifest itself in the mixture with the excitation of quark condensate 
in scalar channel ($\sigma$-meson) is not bad candidate for that role. By the 
way, it could be experimentally observed as a wide resonance.

Summarizing, we understand  our calculation can not pretend to the precise 
quantitative agreement with experimental data and see many things to be 
done. We are planning shortly to consider the problem of instanton profile 
\cite{sim}, to make more realistic description of the PP interaction, 
to push our ansatz beyond the long-length wave approximation analyzing
more precisely 'instanton Jacobian' 
$\left|\frat{\delta A}{\delta \varphi}\right|$.

The authors have benefited from the discussions with many people
but especially with N.O. Agasyan, M.M. Musakhanov, Yu.A. Simonov.  
The paper was accomplished under the Grants of RFFI
97-02-17491 and INTAS (93-0283, 96-0678). 
Two of us (S.V.M. and A.M.S.) acknowledge Prof. M. Namiki and the HUJUKAI Fund for 
the permanent financial support.

\section*{Appendix}
The contribution of the quark determinant to the IL action is given by
the tadpole diagram Eq. (\ref{tdp}) which takes the following form when 
returned to the dimensional variables (see, Eq. (\ref{sc1})) 
$$
\Delta \varphi=\Delta~\frat{(n\kappa)^{1/2}}{\bar\rho^3}~\varphi(0)=
\Delta~ (n\bar\rho^4\kappa)^{1/2}\int d\rho\frat{n(\rho)}{n}
~\int \frat{dz}{\bar\rho^4}~
\frat{\rho(z)-\rho_c}{\bar\rho}~.
$$
Then the IL action, Eq. (\ref{s}), acquires the additional term
$$
\langle S\rangle=
\int d z~n \left\{\langle s\rangle-
\langle \Delta'~\frat{\rho-\rho_c}{\bar\rho}\rangle\right\}~,
$$
where $\Delta'=\frat{4N_c}{n\bar\rho^4}~\mu^2 c(\mu)$
and the mean action per one instanton is given by the following
functional
$\langle s_1 \rangle= \int d\rho~s_1(\rho) n(\rho)/n$ with
$$s_1(\rho)=\beta(\rho)+5 \ln(\Lambda\rho)-\ln \widetilde \beta^{2N_c}+
\beta\xi^2\rho^2n\overline{\rho^2}-\Delta'~(\rho-\rho_c)/\bar\rho~.
$$ 
In order to evaluate the equilibrium parameters of IL we treat the 
maximum principle
$$\langle e^{-S}\rangle\ge\langle e^{-S_0}\rangle
e^{-\langle S-S_0\rangle}$$
adapting it to the simplest version (when the approximating functional
is trivial $S_0=0$). In a sense, this choice of the approximating 
functional should be a little worse than in Ref. \cite{2}. Its only 
advantage comes from the possibility to get the explicit formulae for the 
IL parameters in lieu of solving the complicated transcendental equation.
In equilibrium the instanton size distribution function 
$n(\rho)$ should be dependent on the IL action only, i.e.
$n(\rho)=C e^{-s(\rho)}$ where $C$ is a certain constant. 
This argument corresponds to the maximum principle of Ref. \cite{2}. 
Indeed, if one is going to approach the functional (\ref{s}) as a local form 
$\langle s \rangle= \int d\rho~ s_1(\rho) n(\rho)/n$ 
where
$s_1(\rho)=\beta(\rho)+5 \ln(\Lambda\rho)-\ln \widetilde \beta^{2N_c}+
\beta\xi^2\rho^2n\overline{\rho^2}$,   
it makes the approach  self-consistent. The functional difference 
$\langle s \rangle-\langle s_1 \rangle= \int d\rho~\{ s(\rho)- s_1(\rho)\}
e^{-s(\rho)}/n$ being varied over $s(\rho)$ 
leads then to the result 
$s(\rho)=s_1(\rho)+const$ keeping into the mind an arbitrary
normalization. The maximum principle results in getting the mean action per 
one instanton as the IL parameters function, for instance, average instanton 
size $\bar\rho$. The corrections generated by the 'shifting' terms turn out 
to be small and we consider them in the linear approximation in the deviation 
$\Delta$. The following schematic expansion exhibits how the major 
contribution appears
\begin{equation}
\label{a5}
\langle s_1\rangle=
\frat{\langle (s+\delta)~e^{-s-\delta}\rangle}
{\langle e^{-s-\delta}\rangle}
\simeq
\frat{\langle s~e^{-s}\rangle+
\langle \delta~e^{-s}\rangle}{\langle e^{-s}\rangle}+
\frat{\langle s~e^{-s}\rangle\langle \delta~e^{-s}\rangle-
\langle s~\delta~e^{-s}\rangle \langle e^{-s}\rangle}
{\langle e^{-s}\rangle^2}~,
\end{equation}
here $\delta(\Delta)$ stands for a certain small 'shifting' contribution and 
$s$ is the action generated by the gluon component only. The last term in 
Eq. (\ref{a5}) is small comparing to the first one and we neglect it. 
Then it is clear that evaluating the mean action per one instanton is 
permissible to hold the gluon condensate contribution $s$ only (without the
'shifting' term $\delta$) in the exponential. Hence we have for the mean 
action per one instanton 
$\langle s_1 \rangle= \int d\rho~s_1(\rho) n_0(\rho)/n_0$,
and $n_0(\rho)$ is the distribution function which does not include the
'shifting' term
{\footnote{The 'shifting' term changes the mass of phononlike excitation 
insignificantly. The equilibrium instanton size as dictated by the
condition $\left.\frat{d s(\rho)}{d\rho}\right|_{\rho=\rho_c}=0$ is equal 
then to  $\rho_c=(\alpha+\Delta'~\beta)~\bar\rho,
~\alpha=\left(1-\frat{1}{2\nu}\right)^{1/2},~
\beta=\frat{1}{4\nu}\left\{1-\alpha~\frat{\Gamma(\nu+1/2)}
{\nu^{1/2}\Gamma(\nu)}\right\}$ and the second derivative of action in 
the equilibrium point equals to
  $s^{''}(\rho_c)=\frat{4\nu}{\overline{\rho^2}}+
\frat{2\nu}{\overline{\rho^2}}~\Delta'\left\{\frat{\Gamma(\nu+1/2)}
{\nu^{3/2}\Gamma(\nu)}-\frat{1}{2\nu\alpha}\right\}$. 
Another source of corrections to the kinetic coefficient appears while one 
considers the instanton profile change $A\to A+a$
where the field of correction is 
$a\sim\left.\frat{\partial \rho(x,z)}{\partial x}\right |_{x=z}$. 
This mode could appear  within the superposition ansatz Eq. (\ref{2}) and 
leads 
to  the modifications of quark zero mode ($D(A+a)\psi=0$). 
Fortunately, both corrections to
the kinetic term occurs to be numerically small.}}.
It is possible to obtain for the average squared instanton size and
the IL density that
{\footnote{In order not to overload the formulae with the factors making the
results dimensionless, which are proportional to the powers of $\Lambda$, 
we drop them out hoping it does not lead to the misunderstandings.}}  
\begin{equation}
\label{a7}
r^2 \overline{\rho^2}=\nu~\left\{1+\frat{\Delta'}{r\bar\rho}~
\frat{\Gamma(\nu+1/2)}{2\nu\Gamma(\nu)}\right\}\simeq
\nu~\left\{1+\Delta'~\frat{\Gamma(\nu+1/2)}{2\nu^{3/2}\Gamma(\nu)}\right\}~,
\end{equation}
\begin{equation}
\label{a8}
n=C~C_{N_c}\widetilde \beta^{2N_c}\frat{\Gamma(\nu)}{2r^{2\nu}}~,
\end{equation}
where the parameter $r^2$ equals to
\begin{equation}
\label{a7a}
r^2=\beta \xi^2n\overline{\rho^2}~. 
\end{equation}
Expanding
$\ln\rho=
\ln\bar\rho+\frat{\rho-\bar\rho}{\bar\rho}+
\frat12\frat{(\rho-\bar\rho)^2}{\bar\rho^2}+\cdots$
and using Eq. (\ref{a7}) we can show that
$$\frat{\int d\rho~ n_0(\rho) \ln \rho}{\int d\rho~ n_0(\rho)}=\ln\bar\rho+
\Phi_1(\nu)~,~~
\frat{\int d\rho~ n_0(\rho) \rho}{\int d\rho~ n_0(\rho)}=\bar\rho+
\Phi_2(\nu)~,$$
where $\Phi_1,~\Phi_2$ are the certain function of $\nu$ independent of 
$\bar\rho$. Besides, the average squared instanton size within the precision
accepted obeys the equality
$r^2 \overline{\rho^2}=\Phi(\nu)$,
and $\Phi(\nu)$ is the function of $\nu$ only. Then the mean action per one
instanton looks like
$$\langle s_1 \rangle= -2N_c \ln \widetilde \beta+
(2\nu-1)~\ln\bar\rho+F(\nu)
$$
($F(\nu)$ is again the function of $\nu$ only and its explicit form is
unessential for us here). Calculating its maximum in $\bar\rho$ we receive
$$\bar\rho=\exp\left\{-\frac{2N_c}{2\nu-1}\right\}~,
~\beta=\frat{2bN_c}{2\nu-1}-\ln C_{N_c}~.
$$
From Eqs. (\ref{a7}), (\ref{a7a}) we find the IL density to be as
$$n=\nu~\frat{e^{\frac{8N_c}{2\nu-1}}}{\beta\xi^2}
\left\{1+\Delta'~\frat{\Gamma(\nu+1/2)}{2\nu^{3/2}\Gamma(\nu)}
\right\}~,
$$
and handling Eq. (\ref{a8}) we determine the constant $C$.

The IL parameters occur to be close to the parameter values of the 
Diakonov-Petrov approach \cite{2} and are shown in the following
\begin{center}
Table 3.\\\vspace{0.3cm}
\begin{tabular}{|cccc|cccc|}
\hline
&&D.P.&&&&~M.S.Z.~~&\\\hline
      $N_f$&$\bar\rho\Lambda$&$n/\Lambda^4$  &$\beta$
       &$N_f$&$\bar\rho\Lambda$&$n/\Lambda^4$  &$\beta$\\\hline
0 & 0.37  &0.44&17.48 
&0 & 0.37  &0.44 (0.49)&17.48\\      
1 & 0.30  &0.81&18.86 
&1 & 0.33  &0.63 (0.71)&18.11\\
2 & 0.24  &1.59&20.12  
&2 & 0.28  &1.03 (1.17)&18.91\\
\hline
\end{tabular}
\end{center}
Here $N_f$ is the number of flavours, $N_c=3$, and the IL density at 
$\Delta\ne0$ when the iteration process completed is shown in the 
parenthesis. It is curious to notice the quark influence on the IL
equilibrium state  provokes the increase of the IL density. 
 
In Table 4 we demonstrate the mass gap magnitude $M$ and the wave length in
the 'temporal' direction $\lambda_4=M^{-1}$. To make it more indicative we
show also the average distance between PPs from which is clear, in fact, 
that the adiabatic approximation $\lambda\ge L\sim \bar R>\bar\rho$ is 
valid for the long-length wave excitations of the $\pi$-meson type. 
All the parameters are taken at $\kappa=4\beta$ but the primed ones  
correspond to the kinetic term value of $\kappa=6\beta$. 
\begin{center}
Table 4.\\\vspace{0.3cm}
\begin{tabular}{|clclcl|}
\hline
$N_f$&$M\Lambda^{-1}$&$\lambda\Lambda$
&$M^{'}\Lambda^{-1}$&$\lambda'\Lambda$&$\bar R\Lambda$\\\hline
0&1.21&0.83&0.99&1.01&1.23 (1.2)\\      
1&1.34&0.75&1.09&0.91&1.12 (1.09)\\
2&1.45&0.69&1.18&0.84&0.99 (0.96)\\
\hline
\end{tabular}
\end{center}
Here the parameters in the parenthesis designate the
distance between PP ($\bar R=n^{-1/4}$) when the iteration process is 
completed.

\end{document}